\documentclass[slac_one]{revtex4}
\usepackage{graphicx}
\usepackage{fancyhdr}
\usepackage{units}
\newcommand{\delmsq}[1]{\ensuremath{\Delta m^2_{ #1 }}}

\newcommand{\sinsq}[1]{\ensuremath{\sin^{2}\left(2\theta_{ #1 }\right)}}

\newcommand{\numu}{\ensuremath{\nu_{\mu}}}
\newcommand{\nue}{\ensuremath{\nu_{e}}}
\newcommand{\nutau}{\ensuremath{\nu_{\tau}}}

\newcommand{\minos}{MINOS}

\begin{document}

\title{Neutrino Oscillation Studies with MINOS} %% Paper title goes here


\author{Patricia L. Vahle, for the MINOS Collaboration}
\affiliation{Department of Physics, College of William and Mary, Williamsburg VA  23187}
%
%\author{P. Lucas}
%\affiliation{FNAL, Batavia, IL 60510, USA}

\begin{abstract}
The \minos{} experiment observes the \numu{} beam produced by the
Main Injector accelerator at two detector stations, one \unit[1]{km} and the
other \unit[735]{km} from the neutrino production target.  From the disappearance of neutrinos in flight between the two detectors, we measure the neutrino oscillation parameters $\Delta
m_{23}^2$ and $\sin^2(2\theta_{23})$ and update our previously published
result.  Additionally, we have searched for evidence of additional light
neutrino families which do not couple to the weak interaction.  Finally, we discuss the status of our
search for the appearance of \nue{}, an effect
expected if the mixing angle $\theta_{13}$ is non-zero and required if the
goal of CP violation in the neutrino sector is to be observed.
\end{abstract}

\maketitle

\thispagestyle{fancy}


\section{The MINOS Experiment}
\minos{} is a long baseline neutrino experiment built to study neutrino oscillations~\cite{minoscollab}.  \minos{} employs two magnetized iron tracking/sampling calorimeters to measure the survival probability of a beam of \numu{} produced in the NuMI facility at Fermilab~\cite{Michael:2008bc,Kopp:2005bt}.  NuMI is a conventional two-horn focused neutrino beam with a \unit[675]{m} long decay tunnel.  The horn current and position of the hadron production target relative to the horns can be configured to produce different \numu{} energy spectra.  Data taken in different beam energy configurations is used to tune the Monte Carlo neutrino flux simulation.   The \minos{} Near Detector is installed on-site at Fermilab, \unit[1]{km} downstream of the hadron production target.  The \minos{} Far Detector is \unit[735]{km} downstream from the target in the Soudan Underground Laboratory in Northern Minnesota.  With this baseline and a peak beam energy of \unit[3]{GeV}, \minos{} is sensitive to neutrino oscillations governed by the \delmsq{32} mass splitting.  By design, the two \minos{} detectors respond nearly identically to neutrino interactions.  All \minos{} analyses measure the beam flavor composition and energy spectrum in the Near Detector and use that data to predict the Far Detector flavor composition and spectrum in the absence of oscillations.  The Far Detector measurement is then compared to the non-oscillated prediction based on the Near Detector data.  The use of two detectors mitigates most systematic uncertainties associated with the measurement of the oscillation parameters.  For instance, systematic errors arising from neutrino flux mismodeling and neutrino interaction uncertainties are almost negligible in the two detector experiment.

The \minos{} oscillation physics goals are threefold.  The flagship measurement involves determining the disappearance probability of \numu{} events as a function of energy.  The energy dependence of disappearance of \numu{} type neutrinos provides a precision measurement of the oscillation parameters governing this mode of oscillations, namely $|\delmsq{32}|$ and \sinsq{23}, and enables discrimination between the oscillation hypothesis and alternative models proposed to explain the disappearance.  \minos{} also measures the rate of neutral current (NC) interactions in the Far Detector relative to the the Near.  In standard three flavor oscillation scenarios, the NC rate should not be affected by oscillations; any reduction in the rate of NC interactions or distortion in the energy spectrum observed in the Far, relative to the Near, could be evidence for a sterile neutrino.  Finally, \minos{} is sensitive to the subdominant $\numu{}\rightarrow\nue{}$ oscillation mode.  An excess of \nue{} in the Far Detector, again relative to the Near, could measure, or further limit, \sinsq{13}.

\section{\numu{} Disappearance}
\begin{figure*}[t]
\centering
\begin{minipage}[l]{0.49\textwidth}
\includegraphics[width=\textwidth]{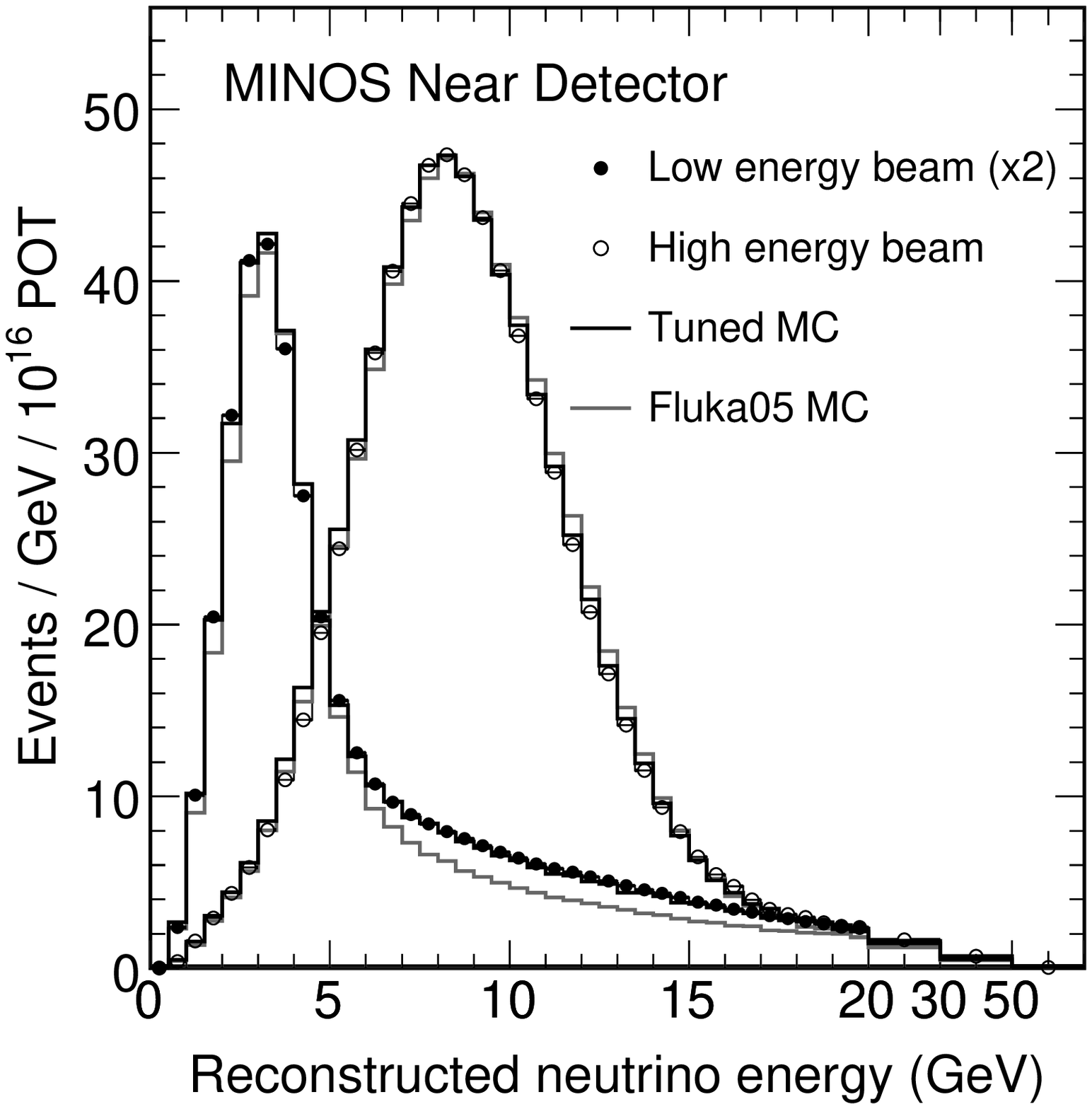}
\end{minipage}
\begin{minipage}[l]{0.49\textwidth}
\includegraphics[width=\textwidth]{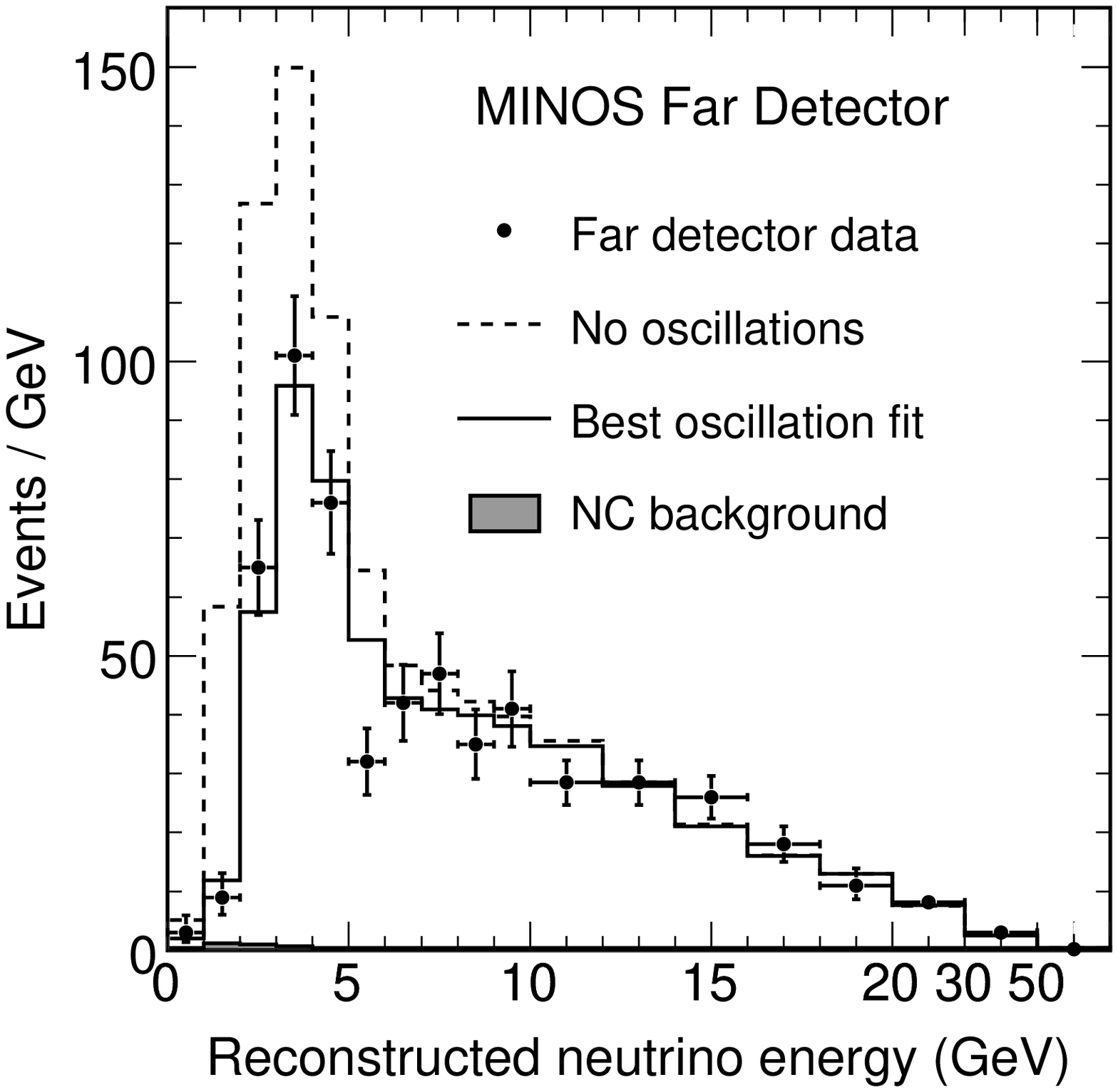}
\end{minipage}
\caption{(Left) Energy spectra in the \minos{} Near Detector for two different beam energy configurations used in the oscillation analysis, compared to the Monte Carlo simulation.  (Right) Comparisons of the Far Detector data (different beam energy configurations combined) with the prediction of the spectrum with and without the effect of oscillations.} 
\label{fig:ccspec}
\end{figure*}

\minos{} probes the \numu{} disappearance probability by looking for an energy dependent suppression of \numu{} charged current (CC) events in the Far Detector relative to the prediction of the spectrum in the case of no oscillations based on the Near Detector measurement.  Charged current \numu{} interactions are selected in the \minos{} detectors by first identifying a track in an event, then using a multivariate likelihood algorithm based on variables that characterize the track as being muon-like. 
% The variables used in the characterization are event length, the average pulse height per plane along the track, the fluctuation of the energy deposited along the track, and the transverse energy deposition profile along the track.  With this algorithm, \numu{} CC events are identified with 81.5\% efficiency;  the main background is a 0.6\% contamination of NC events~\cite{Adamson:2008zt}.  
The left panel of Figure~\ref{fig:ccspec} shows the spectrum of selected events in the Near Detector compared to the Monte Carlo simulation.  Two NuMI beam energy configurations are shown in the plot.  The low energy tune enhances the number of \numu{} at low energies where oscillations are expected to be maximal.  The high energy tune augments the \numu{} flux at high energies where oscillations are not expected to occur.  The extra events at the higher energy help constrain the normalization in the Near to Far prediction and also aid in the discrimination among alternate models of \numu{} disappearance.  Data from both of these beam energy configurations was used in the most recent oscillation analysis, corresponding to a total exposure of \unit[3.36$\times 10^{20}$]{protons-on-target}.

\begin{figure*}[t]
\centering
\begin{minipage}[l]{0.49\textwidth}
\includegraphics[width=\textwidth]{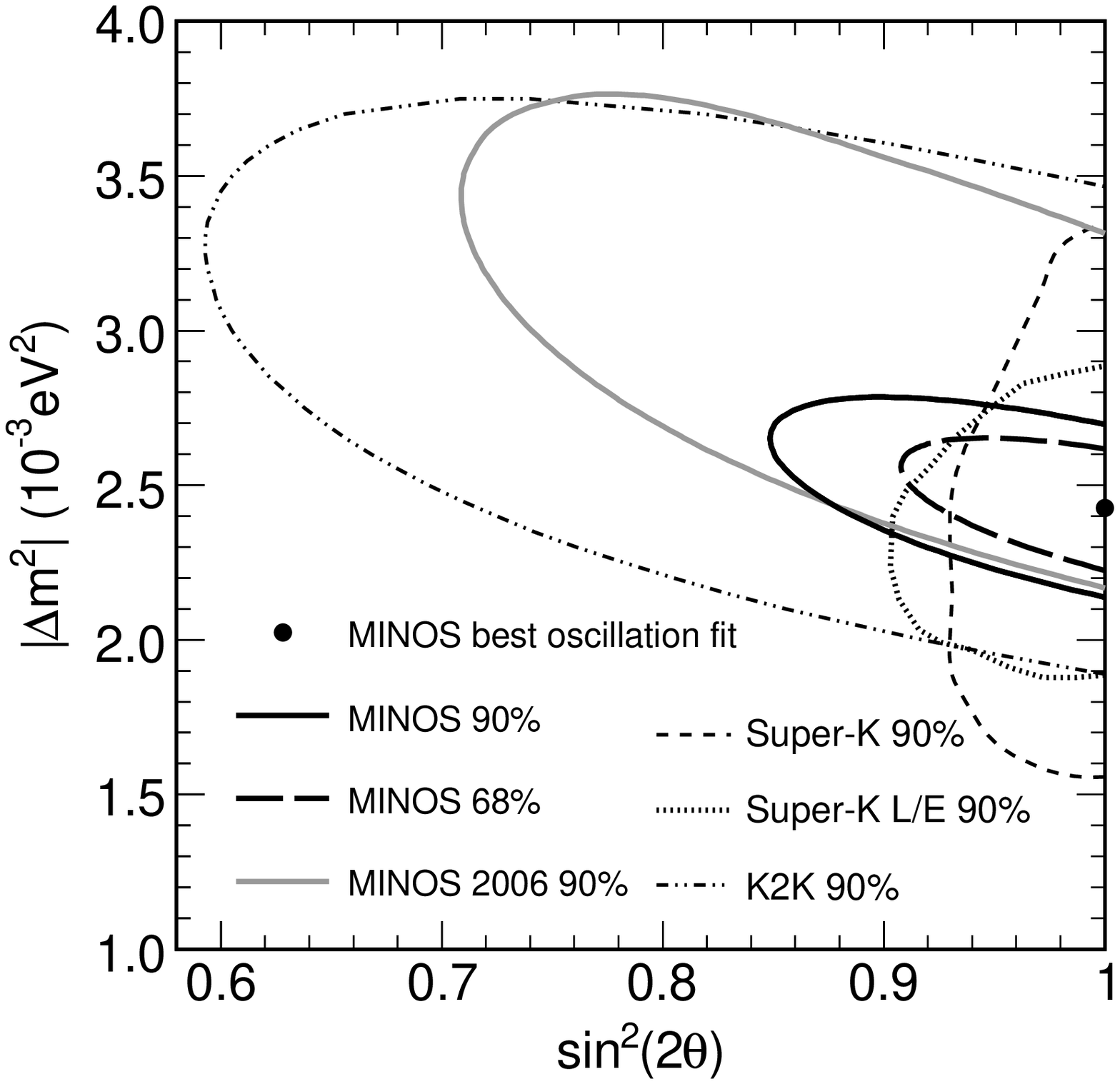}
\end{minipage}
\begin{minipage}[l]{0.49\textwidth}
\includegraphics[width=\textwidth]{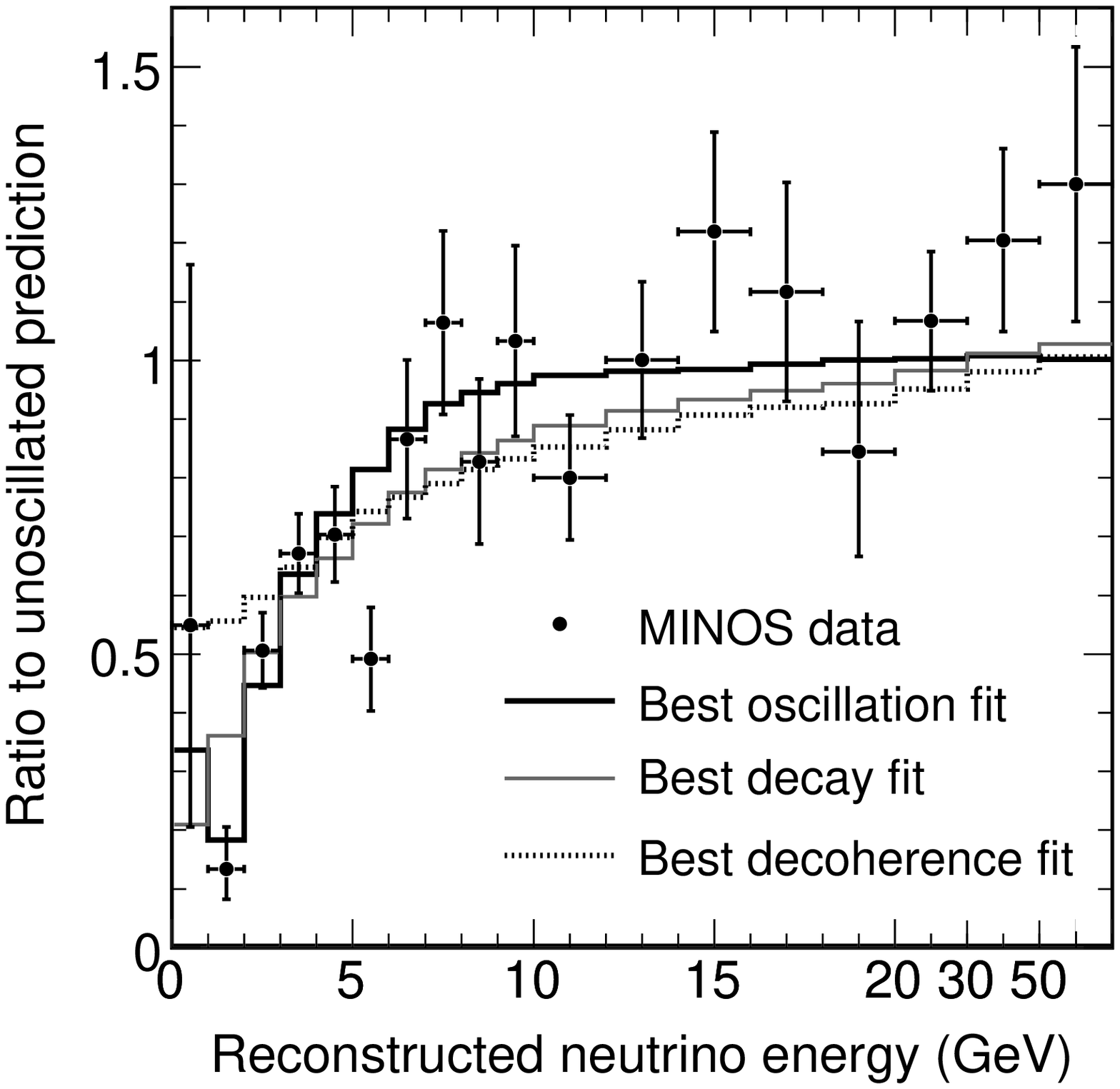}
\end{minipage}
\caption{(Left) Contours for the oscillation fit including systematic errors.  Also shown are results from previous experiments~\cite{Ashie:2004mr,Ahn:2006zz} and the earlier \minos{} result~\cite{prl2006,Adamson:2007gu}.  (Right) Ratio of the Far Detector data and the expected spectrum in the absence of oscillations.  Also shown are the best fit to the oscillation hypothesis and alternative models of neutrino disappearance.} 
\label{fig:cccontour}
\end{figure*}

%When the same event selection criteria are applied to the 
In the Far Detector data, we observe 848 events in the Far Detector over the full energy range of the NuMI beam.  The unoscillated expectation based on the Near Detector data is 1065$\pm$60 (syst.)~\cite{Adamson:2008zt}.  The spectrum of the Far Detector events are shown to the right in Figure~\ref{fig:ccspec}.  The deficit of events can be clearly seen at energies below \unit[5]{GeV}.  Under the assumption that the deficit is caused by $\numu\rightarrow\nutau{}$ oscillations, the parameters $|\delmsq{}|$ and \sinsq{} are extracted from a fit of the oscillation probability to the data:

\begin{equation}
P(\numu{}\rightarrow\numu{})=1-\sinsq{}\sin^{2}\left(1.27\delmsq{}\frac{L}{E}\right)
\end{equation}

\noindent where L[km] is the distance from the target, E[GeV] is the neutrino energy, and $|\delmsq{}|$ is measured in ${\rm eV^{2}}$.  The best fit parameters are those which maximize a likelihood that includes penalty terms for the three dominant systematic uncertainties: a 50\% uncertainty in the NC contamination; a 10.3\% uncertainty in the absolute hadronic energy scale; and a 4\% error on the Far Detector predicted event rate, arising from the combination of uncertainties in the detectors' fiducial mass, event identification, and proton-on-target counting.  The resulting best fit is shown in Figure~\ref{fig:ccspec}.  The values obtained for the oscillation parameters, when the fit is constrained to lie in the physical region, are $|\delmsq{}|=(2.43\pm0.13)\times 10^{-3}~{\rm eV^{2}}$ and $\sinsq{}>0.95$ at the 68\% C.L~\cite{Adamson:2008zt}.  The $\chi^{2}$ of the fit is 90 for 97 degrees of freedom.  The 68\% and 90\% C.L. intervals for the oscillation parameters are shown to the left in Figure~\ref{fig:cccontour}.

The data has also been compared to other models proposed to explain the disappearance of neutrinos in flight, specifically the decay of neutrinos into lighter particles~\cite{Barger:1998xk} and the decoherence of the neutrino's quantum mechanical wave packet~\cite{Fogli:2003th}.  The right panel of Figure~\ref{fig:cccontour} shows the ratio of the Far Detector data to the unoscillated prediction.  It also shows the ratio predicted in the case of oscillations with the best fit parameters and the predictions of the decay and decoherence models.  The \minos{} data disfavor decay at the 3.7$\sigma$ level and decoherence at the 5.7$\sigma$ level relative to the oscillation hypothesis~\cite{Adamson:2008zt}.  

\section{Neutral Current Event Rate}
In a standard three-neutrino oscillation scenario, the NC event rate should not be distorted by oscillations.  Any changes in the NC energy spectrum could be interpreted as evidence for mixing into a sterile neutrino flavor.  \minos{} sees no evidence of such distortions.  The Far Detector NC energy spectrum is shown to the left in Figure~\ref{fig:ncspec}.  With this data, \minos{} limits the fraction of neutrinos that transform into the sterile state to less than 68\% at the 90\% C.L., in the case of no \nue{} appearance~\cite{Adamson:2008jh}.  Any neutrinos that transform into the electron flavor state on the way to the Far Detector would directly affect this result.  If \nue{} appearance occurs at the currently allowed maximal rate, the \minos{} data limit mixing to the sterile state at less than 80\% at the 90\% C.L.~\cite{Adamson:2008jh}.

\section{\nue{} Appearance}
\begin{figure*}[t]
\centering
\begin{minipage}[l]{0.49\textwidth}
\includegraphics[width=\textwidth]{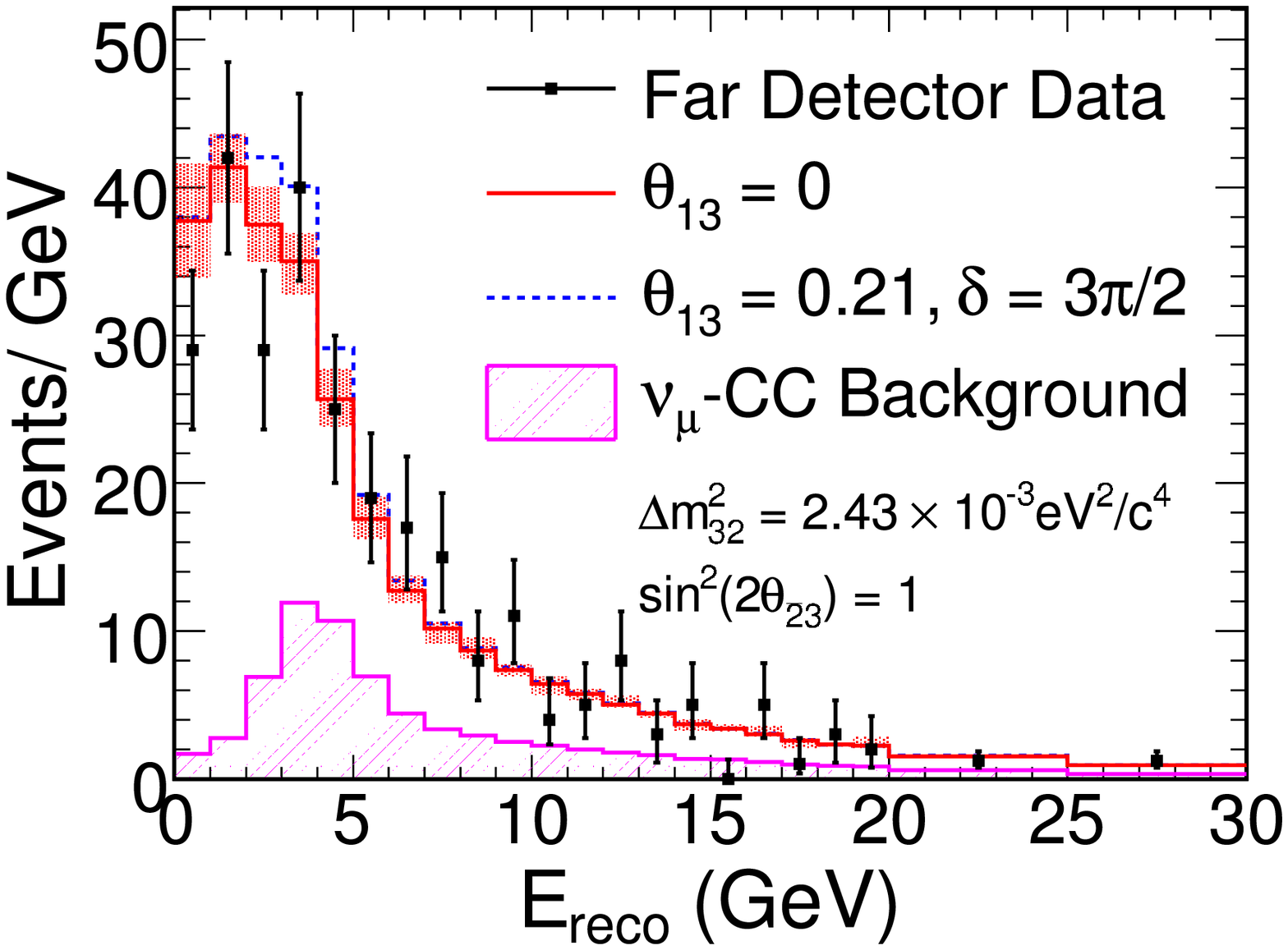}
\end{minipage}
\begin{minipage}[l]{0.49\textwidth}
\includegraphics[width=\textwidth]{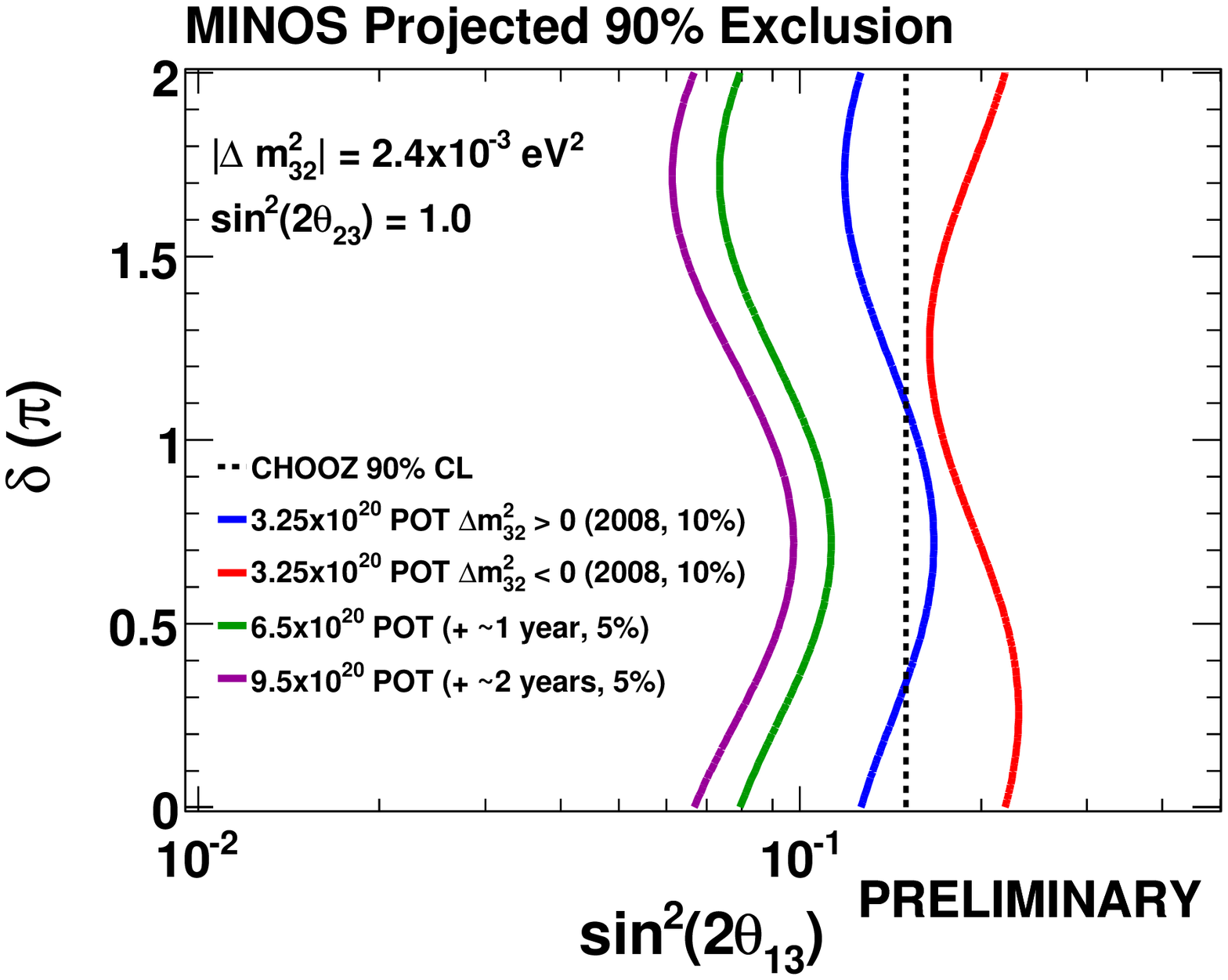}
\end{minipage}
\caption{(Left) Spectrum of observed NC like events in the Far Detector with predictions for the standard neutrino oscillation hypotheses with maximal and no \nue{} appearance.  Systematic uncertainties are shown as shaded bands on the predicted spectrum.  (Right) The projected sensitivity of MINOS to \sinsq{13} as a function of $\delta_{CP}$ .  The right most curves show the 90\% confidence level upper limit on \sinsq{13} if no excess events are observed at the current exposure for both the normal and inverted hierarchies.  Other curves show the future MINOS projected sensitivity for the normal hierarchy with higher exposures and a reduced systematic error.  The dotted black line shows the upper limit set by the CHOOZ experiment~\cite{Apollonio:2002gd}. } 
\label{fig:ncspec}
\end{figure*}

Finally, beyond investigating $\numu{}$ disappearance, MINOS is sensitive to the possible $\numu{}\rightarrow\nue{}$ oscillation mode.  The signature of such a mode would be an excess of \nue{} events in the \numu{} beam in the Far Detector relative to the measurement of backgrounds in the Near Detector.  Being an appearance search and because of the long baseline, the \minos{} results are dependent on both the CP-violating phase $\delta_{CP}$ and the mass hierarchy.  While the first analysis of this mode of oscillation is still underway within the MINOS collaboration, The right panel of Figure~\ref{fig:ncspec} shows the sensitivity MINOS expects to achieve in the present and future analyses.  With the data already collected, MINOS can further limit the value of \sinsq{13}, or with luck, be the first experiment to detect this mode of oscillation.

\section{Summary}
The \minos{} experiment is currently exploring many facets of the neutrino oscillation problem.  Increasing exposures promise increasingly precise measurements of $|\delmsq{}|$ and \sinsq{23} and increased confidence in the oscillation description relative to other models.  \minos{} neutral current event rate studies show no evidence of distortions that would be characteristic of mixing to sterile neutrinos.  Finally, \minos{} is sensitive to $\numu{} \rightarrow \nue{}$ transitions.  A first analysis of the \nue{} event rate in the \minos{} Far Detector will further limit \sinsq{13}, or even provide the first direct experimental evidence of this mode of oscillation.

\bibliographystyle{h-physrev3}
%\begin{thebibliography}{9}   % Use for  1-9  references
%\begin{thebibliography}{99} % Use for 10-99 referencesXB

\bibliography{minos}

%\end{thebibliography}

\end{document}